\documentclass{article}

\usepackage{verbatim}
\usepackage{amsmath}
\usepackage{amsfonts}
\usepackage{amssymb}
\usepackage[hidelinks]{hyperref}
\usepackage[all]{hypcap}

\title{
	A Generalization of the Minisum and Minimax Voting Methods \\
}
\author{Shankar N. Sivarajan\\
	Undergraduate, Department of Physics\\
	Indian Institute of Science\\
	Bangalore 560 012, India\\
\tt{shankarnsivarajan@gmail.com}\\
\\
Faculty Advisor: Prof. Y. Narahari\\
Department of Computer Science and Automation
}

\date{}

\usepackage{titlesec}
\renewcommand\thesection{\Roman{section}}
\titleformat{\section}[block]{\Large\scshape\centering}{\thesection.}{1em}{}

\begin{document}

\maketitle
	
\begin{abstract} \noindent
In this paper, we propose a family of approval voting-schemes for electing committees based on the preferences of voters. In our schemes, we calculate the vector of distances of the possible committees from each of the ballots and, for a given $ p $-norm, choose the one that minimizes the magnitude of the distance vector under that norm. The minisum and minimax methods suggested by previous authors and analyzed extensively in the literature naturally appear as special cases corresponding to $ p = 1 $ and $ p = \infty, $ respectively. Supported by examples, we suggest that using a small value of $ p, $ such as 2 or 3, provides a good compromise between the minisum and minimax voting methods with regard to the weightage given to approvals and disapprovals. For large but finite $ p, $ our method reduces to finding the committee that covers the maximum number of voters, and this is far superior to the minimax method which is prone to ties. We also discuss extensions of our methods to ternary voting.
\end{abstract}

\section{Introduction}

In this paper, we consider the problem of selecting a committee of $ k $ members out of $ n $ candidates based on preferences expressed by $ m $ voters. The most common way of conducting this election is to allow each voter to select his favorite candidate and vote for him/her, and we select the $ k $ candidates with the most number of votes. While this system is easy to understand and implement, upon scrutiny, there arise certain unfavorable aspects.

Suppose we wish to elect a committee of size  $ k = 1,$  and that there are two candidates, a Conservative $ A, $ and a Liberal $ B, $ with $ B $ expected to win.  In this case, it is to candidate $ A $'s benefit to introduce a candidate $  C $ whose ideology is similar to that of candidate $ B $ in order to draw votes away from $B$, thereby ensuring his own victory. This may be accomplished in practice by providing campaign funding, for example. This is illustrated by the following example. Assume the preferences of the voters for the various candidates are as indicated in Table~\ref{tab:all preferences}. The first entry $ (A, B, C) =30\%$ means that 30\% of the voters rank the candidates in decreasing order of preference as  $ A  $ followed by $ B, $ and then by $C$. As a result of these preferences, in an election between candidates $ A $ and $ B $ alone, $ B $ would win 56\% of the votes.  But in a three-way election, $ A $ would win, with the votes distributed as in Table~\ref{tab:one preference}.  Similar problems  can arise in the selection of committees with $ k > 1 $ as well.

\begin{table}[ht]
	\begin{center}
		\begin{tabular}{|c|c|}
			\hline
			Preference vector & Percentage of votes \\ \hline
			$ (A, B, C) $   &         30          \\ \hline
			$ (A, C, B) $   &         10          \\ \hline
			$ (B, A, C) $   &          6          \\ \hline
			$ (B, C, A) $   &         25          \\ \hline
			$ (C, A, B) $   &          4          \\ \hline
			$ (C, B, A) $   &         25          \\ \hline
		\end{tabular} 
		
		\caption{Example of a ballot with full preference list indicated.}
		\label{tab:all preferences}
		
	\end{center}
\end{table}

\begin{table}[ht]
	\begin{center}
		
		\begin{tabular}{|c|c|}
			\hline Candidate & Percentage of votes \\ 
			\hline $ A $ & 40 \\ 
			\hline $ B $ & 31 \\ 
			\hline $ C $ & 29 \\ 
			\hline 
		\end{tabular} %Insert example of Condorcet criterion, with Liberals and Conservatives.
		
		\caption{The ballot from Table~\ref{tab:all preferences} with only the top preference indicated.}
		\label{tab:one preference}
		
	\end{center}
\end{table}

%Table with ballors with preferences (A, B, C) , (A, C, B) etc.
 
One way to counteract this is to conduct preliminary elections within the party, in our case the Liberals, to select candidate $ B $ to represent them, and then conducting an election between candidates $ A $ and $ B$. But we seek methods to make a decision based on a single election.

The reason candidate $  B $ is preferable to $ A $ is that if we compare the two of them, a larger percentage of the voters would prefer $ B. $ In fact, if we compare any two candidates, $ B $ is preferred. This type of candidate is called a \emph{Condorcet winner,} and this criterion, that she be preferred by a majority in any pairwise comparison is called the \emph{Condorcet criterion.} However, every election need not have a Condorcet winner. 
A voting method that uses the Condorcet criterion to select a winner is called a \emph{Condorcet method}.

To check whether the Condorcet criterion is satisfied, we may ask each voter to supply his/her full list of preferences, as in Table~\ref{tab:all preferences}, and if there is a Condorcet winner, we can elect her to the committee. However, if there is no Condorcet winner, we still need a method to choose the winner.  We can compare the candidates pairwise and choose the candidate who wins the most pairwise comparisons, and this is called the Copeland Method. 
%However, if there is no Condorcet winner, this method will usually result in a tie.

To elect a single winner, we can also use \emph{approval voting}~\cite{approval}. We ask each voter to approve two candidates, and we select the candidate with the most number of approvals. In our example, assuming that each voter approves of his top two preferences, $ A $ is approved by 50\% of the voters, $ B $ is approved by 86\%, and $ C $, by 64\%. This elects candidate $  B, $ as desired.

\begin{comment}
In general, we would like to construct a voting rule which satisfies the following  basic ``fairness'' criteria. 

\begin{itemize}
	\item If every voter prefers candidate $ A $ over $ B, $ the group also prefers candidate $ A $ over $ B. $
	\item If every voter's preference between $ A $ and $ B $ remains unchanged, then the group's preference between $ A $ and $ B $ will also remain unchanged, even if voters' preferences between other pairs change.
	\item Non-dictatorship: There is no single voter who possesses the power to determine the group's preference.
\end{itemize}

While these conditions all seem reasonable, Arrow proved that there is no voting rule that satisfies all of them simultaneously~\cite{arrow, arrow_proof}. This is known as \emph{Arrow's Impossibility Theorem}.
\end{comment}

Approval voting is not a Condorcet method, in general. However, in large elections with, say, a million voters, and about twenty candidates, it is impractical to require the voters to provide a full list of preferences, and hence Condercet or Copeland methods are infeasible but approval voting is eminently feasible. Motivated by this feature, and also because it is amenable to the analytical methods we employ, we consider only approval voting and some of its variants in this paper.
 
\section{Shortcomings of the Minisum and Minimax Voting Methods}

Suppose we have to elect a committee of $ k $ representatives from a pool of $ n $ candidates. Further, suppose there are $  m $ voters and each of them can approve of  an arbitrary number of  candidates. 

We will consider the example used in~\cite{minimax_AAAI} shown in Table~\ref{tab:paper example}.  In this case $ n = 5, m = 4 $ and $ k = 2. $  We represent the ballot of voter $1$ by the $ n $-dimensional binary vector $ (1, 1, 0, 0, 1), $ and similarly for all the other voters.

%The candidates are represented by letters, and the voters by italicized numbers.  

\begin{table}[ht]
	\begin{center}
		\begin{tabular}{|c|c|c|c|c|c|}
			\hline
			  & $ A $ & $ B $ & $ C $ & $ D $ & $ E $ \\ \hline
			\emph{1} &   1   &   1   &   0   &   0   &   1   \\ \hline
			\emph{2} &   1   &   1   &   0   &   0   &   0   \\ \hline
			\emph{3} &   1   &   1   &   1   &   1   &   0   \\ \hline
			\emph{4} &   0   &   0   &   1   &   1   &   1   \\ \hline
		\end{tabular}
		
		\caption{The example from~\cite{minimax_AAAI} showing the approvals of the voters \emph{1}, \emph{2} \ldots \emph{4} for the candidates $  A, $  $ B $ \ldots $E$.  An approval is indicated by a ``1''.}
		\label{tab:paper example}
	
	\end{center}
\end{table} 

Now, we wish to select a committee of size $ k = 2, $ and the most common method is called the \emph{minisum} method.  This means we simply select the top $ k $ candidates with the most votes, in this case $  A $ and $  B. $  Like ballots, we also represent committees by binary vectors of length $n$ with a $1$ denoting membership in the committee. Thus we represent the committee consisting of $A$ and $B$ by the binary vector $  (1, 1, 0, 0, 0). $ This method is called the minisum strategy because, if you take the Hamming distances of this committee-vector to each of the $ m $ ballot-vectors, in this case 1, 0, 2, 5, and add them, this is the committee vector that minimizes this sum.  We now prove this result. 

Let the candidates be numbered from 1 to $n$, and let the number of approvals for them be $ a_1, a_2, \ldots a_n $, respectively. Without loss of generality, assume $ a_1 \geq a_2 \geq \ldots \geq a_n$. Let the committee that minimizes the sum of Hamming distances be $ (i_1, i_2, \ldots i_k). $  
This minimum value is equal to \[ (a_1 + a_2 + \ldots + a_n) - (a_{i_1} + a_{i_2} + \ldots a_{i_k}), \] which, neglecting ties for the $ k^\textrm{th} $ position,  is obviously minimized when \[ (i_1, i_2, \ldots i_k) = (1, 2, \ldots k). \]

This result allows us to consider the problem of finding the minisum committee as an optimization problem since we have a computable number, the sum of Hamming distances, that enables us to compare any two committees and choose the better one.
But this method does not weigh disapprovals enough (as seen earlier in Table~\ref{tab:one preference}) and so, a method called the \emph{minimax} was introduced~\cite{minimax_approx}. This method chooses the committee which minimizes the maximum Hamming distance to the ballots. Under this, we elect the committee of $ A $ and $ C $, i.e., the vector $  (1, 0, 1, 0, 0). $ 

We will demonstrate the inadequacies of these methods bv considering the example in Table~\ref{tab:main example}, with $ m = 1000 $ voters to elect a  committee of $  k = 2 $ members. We name the $ n = 4 $ candidates $ A_1, A_2, B_1 $ and $ B_2 $ to indicate the correlation of ballots between candidates $ A_1 $ and $ A_2 $ and between $ B_1 $ and $ B_2. $ For simplicity, in the rest of the paper, we restrict each of the voters to approve the same number of candidates as the size of the committee, in this case, two. 

\begin{table}[ht]
	\begin{center}
		\begin{tabular}{|c|c|}
			\hline
			Candidate pair & Votes \\ \hline
			$ \{A_1, A_2\} $ &  500  \\ \hline
			$ \{A_1, B_1\} $ &  100  \\ \hline
			$ \{A_1, B_2 \}$ &  10   \\ \hline
			$ \{A_2, B_1\}$ &  20   \\ \hline
			$ \{A_2, B_2\} $ &  20   \\ \hline
			$ \{B_1, B_2\} $ &  350  \\ \hline
		\end{tabular} 
		\caption{An example of a ballot which exhibits correlations of votes, possibly between candidates with the same ideology.}
		\label{tab:main example}
	\end{center}
\end{table}

 Using the minisum method, we would elect candidates $ A_1 $ and $ A_2 $, as they have the most number of approvals, as shown in Table~\ref{tab:votes per candidate}. However, this leaves 35\% of the voters without either of their choices in the committee. 
 
 \begin{table}[ht]
 	\begin{center}
 		\begin{tabular}{|c|l|}
 			\hline
 			Candidate  & Number of approvals  \\ \hline
 			$ A_1 $    & 500 + 100 + 10 = 610 \\ \hline
 			$ A_2 $    & 500 + 20 + 20 = 540  \\ \hline
 			$ B_1 $    & 100 + 20 + 350 = 470 \\ \hline
 			$ B_2 $    & 10  + 20 + 350 = 380 \\ \hline
 		\end{tabular} 
 		\caption{The number of approvals per candidate, using the ballots from Table~\ref{tab:main example}.}
 		\label{tab:votes per candidate}
 	\end{center}
 \end{table}
 
 The minimax method is even worse because it does not narrow the choices at all; all of the $\binom{n}{k} = \binom{4}{2} = 6 $ committees yield the same minimax distance and it results in all ballots tied at a Hamming distance of 4. Such tie is almost certain whenever the number of voters is large but the number of possible committees is small.  For example, when there are a million voters and a committee of size $2$ to be chosen from $10$ candidates, it is highly likely that there is one of the million voters who is against both candidates in every possible one of the $\binom{10}{2} = 45$ committees. In this case, the minimax method will yield the same minimax distance of $4$ for every one of the $45$ possible committees. Related papers~\cite{minimax_comm, minimax_AAAI} do not elaborate upon this shortcoming but instead focus on efficient computation of the minimax solution.
  
Another problem with the minimax method is that it weighs disapprovals too highly. A single voter can change the outcome of the entire election. Suppose 1000 voters are choosing a committee of size two from four candidates, $\{A_1,A_2,B_1,B_2\}$. Further suppose that the first 998 vote for ${A_1, A_2}$, the 999th voter for $\{A_1,B_2\}$ and the 1000th voter for $\{B_1,B_2\}$. Then the maximum distance to $\{A_1,A_2\}$ is $4$ due to the 1000th voter's ballot, but the maximum distance to $\{A_1,B_2\}$ is only two, and therefore the minimax method elects the committee $\{A_1,B_2\}$. Candidate $A_2$ who was approved by 998 of the 1000 voters is not in the committee! 

An intuitively superior choice for the committee based on the ballots in Table~\ref{tab:main example} would be $ \{A_1, B_1\}, $ where 98\% of the voters get at least one of their choices on the committee. This seems fairer because a committee where I get one choice and you get one choice should be preferred over another committee where one of us gets two choices, and the other none. The minisum method does not prefer one of these committees over the other but we will now demonstrate a method that does so, and chooses the committee $\{A_1, B_1\}$ in this example.

\section{The $p$-norm Minimization Method}

We represent each of the possible committees as a binary vector $ (c_1, c_2, \ldots, c_n)$ with $ c_i = 1 $ if candidate $ i $ is in the committee. Each of the $ m $ approval ballots is a vector of the same form, $(b_1, b_2, \ldots,b_n)$. For each candidate committee-vector, we calculate the vector of Hamming distances of the ballot-vectors from it. This vector has size $ m, $ the number of voters. The components of this vector of Hamming distances are all even (This is a property of binary vectors of equal 1-norm.)  In the case of the committee $ \{A_1, A_2 \}, $  this vector will have a 1000 components, with 500 of these being 0, 150 of these being 2, and 350 of these being 4. For each of the candidate committee-vectors,  this distance information is summarized in Table~\ref{tab:distances}.

\begin{table}[ht]
	\begin{center}
		\begin{tabular}{|c|c|c|c|}
			\hline Committee & \# of ballots & \# of ballots & \# of ballots \\ 
			\hline & at distance 0 & at distance 2 & at distance 4 \\
			\hline $ \{A_1, A_2\} $ & 500&  150 & 350 \\ 
			\hline $ \{A_1, B_1\} $ & 100 & 880 & 20 \\ 
			\hline $ \{A_1, B_2\} $ & 10 & 970 & 20 \\ 
			\hline $ \{A_2, B_1\} $ & 20 & 970 & 10 \\ 
			\hline $ \{A_2, B_2\} $ & 20 & 880 & 100 \\ 
			\hline $ \{B_1, B_2\} $ & 350 & 150 & 500 \\ 
			\hline 
		\end{tabular} 
		\caption{Number of ballots at different distances from the possible committees, using the ballots from Table~\ref{tab:main example}.}
		\label{tab:distances}
	\end{center}
\end{table}

Suppose for a certain committee, the number of ballots at a Hamming distance of $ i $ is $ \nu_i . $ The $ p $-norm of the Hamming distance vector for this committee is, therefore, 
\[ \left( \sum_i \nu_i \cdot i^p  \right)^{1/p}. \] 
We then find the committee that minimizes this $ p $-norm. 

We investigate the result of using intermediate values of $ p $ in Table~\ref{tab:p norm distances}. Minimizing the 1-norm yields the same solution as minisum; minimizing the $ \infty $-norm yields the same solution as minimax. Notice that as the value of $ p \to \infty, $ the value of the distance tends to 4, the maximum Hamming distance. 

\begin{table}[htb]
	\begin{center}
		\begin{tabular}{|*{7}{c|}}
			\hline
			               &                      \multicolumn{6}{c|}{Value of $ p $ in the $ p $-norm}                       \\ \hline
			  Committee    &       1       &       2        &       3        &       4        &      10       &      100      \\ \hline
			$ \{A_1, A_2\} $ & \textbf{1700} &     78.74      &     28.68      &     17.42      &     7.19      &     4.24      \\ \hline
			$ \{A_1, B_1\} $ &     1840      & \textbf{61.97} & \textbf{20.26} &     11.77      &     5.42      &     4.12      \\ \hline
			$ \{A_1, B_2\} $ &     2020      &     64.81      &     20.83      &     11.99      &     5.42      &     4.12      \\ \hline
			$ \{A_2, B_1\} $ &     1980      &     63.56      &     20.33      & \textbf{11.60} & \textbf{5.08} & \textbf{4.09} \\ \hline
			$ \{A_2, B_2\} $ &     2160      &     71.55      &     23.78      &     14.11      &     6.34      &     4.19      \\ \hline
			$ \{B_1, B_2\} $ &     2300      &     92.74      &     32.14      &     19.00      &     7.45      &     4.26      \\ \hline
		\end{tabular} 
		\caption{$ p $-norm of the $ m $-vector for all the possible committees for different values of $ p,$ using the ballots from Table~\ref{tab:main example}. The value in bold is the minimizer for its column.}
		\label{tab:p norm distances}
	\end{center}
\end{table}

At higher values of $p$, the number of voters who have not voted for a given committee is weighted higher, compared to the number of voters who have voted for at least one member of the committee.  We suggest that using this method, choosing a small value of $p$ such as 2 or 3, is a good compromise between the minisum and minimax solution, since it weighs disapprovals more than minisum, but is much less likely to tie than minimax.\footnote[1]{The $p$-norm method will produce a tie for finite $p$, if the number of ballots at each distance is the same for two candidate committees. In this case, there will be a tie between these two committees for all values of $p$. This is much less likely than a tie in the minimax case where two committees will be tied if there one ballot at the maximum possible distance from each of them.}

Note that for $ p, $ equal to 2 or 3, this method chooses the committee $ \{A_1, B_1\}, $ which contains at least one choice of 98\% of the voters. This is the intuitively superior choice we identified earlier.

It is clear that this method generalizes for larger values of $ k, $ the committee size, with the distances ranging across all the even integers from 0 to $ 2k $, assuming $n \ge 2 k$. 

\subsection*{Maximum Coverage Problem (\boldmath$ p \to \infty $)}

In this example, note that the optimal solution for $ p \geq 3 $  always produces the committee $\{A_2, B_1\}.$ Observe that this is the committee with the least number of voters who did not approve of any member in it, i.e., 1\%. It is easy to show this is true in general: For sufficiently large but finite $p$, the $p$-norm method produces the solution with the least number of voters at distance $ 2k, $ and therefore the committee with the ``maximum coverage''. Intuitively, this method corresponds to weighing disapprovals more than approvals.

Let the committees $\mathcal{ C }$ and $\mathcal{ C^\prime} $ have $ \nu_i $ and $ \nu^\prime_i $ voters respectively at a Hamming distance of $ i. $ Let the $ p $-norm of the vector of Hamming distances from $\mathcal{C}$ be $ ||\mathcal{C}||_p. $ Since 
$$
\lim_{p \to \infty} \frac{||\mathcal{C}||_p}{||\mathcal{C^\prime}||_p} 
= \lim_{p \to \infty} \frac{\left( \sum\limits_{i = 0}^{2k} \nu_i \cdot i^p  \right)^{1/p}}{\left( \sum\limits_{i = 0}^{2k} \nu^\prime_i \cdot i^p  \right)^{1/p} }
= \lim_{p \to \infty} \frac{\left( \sum\limits_{i = 0}^{2k} \nu_i \cdot \left( \frac{i}{2k} \right)^p  \right)^{1/p}}{\left( \sum\limits_{i = 0}^{2k} \nu^\prime_i \cdot \left( \frac{i}{2k} \right)^p  \right)^{1/p} } 
=  \left( \frac{\nu_{2k}}{\nu^\prime_{2k}} \right)^{1/p},
$$
we see that for sufficiently large $p$, the committee with minimum $p$-norm is the one which has the minimum number of voters at distance $2k$. This committee minimizes the number of voters who disapprove of every one of its members, and thereby  maximizes the number of voters who approve of at least one of its members. We can intuitively think of the Maximum Cover method as minimizing disapprovals, and in that sense, the other extreme of minisum, which maximizes approvals.

The maximum coverage problem~\cite{coverage_general} is a classic question in computer science and computational complexity theory, and in its standard form, it is phrased as follows:

\begin{quote}
	You are given several overlapping sets and a number $ k. $ You must choose at most $ k $ of the sets such that the union of the selected sets is of maximum size.
\end{quote}

The voting problem can be interpreted as an adaptation of this.  The sets we are given are the voters who approved of a particular candidate. We are to choose $k$ candidates in order to maximize the number of voters with at least one of their approved candidates on the committee. We will treat our example in Table~\ref{tab:main example} as an instance of this problem.

It is easy to see that the maximum cover in this case will be the committee $ \{A_2, B_1\},$ covering 99\% of the voters. This is the same result when $p$ is large but finite. While it is known that the maximum coverage problem is NP-hard, if the committee size $k$ is small, the number of possible committees, $\binom{n}{k}$ is also modest, and the NP-hardness of the maximum cover problem does not pose a computational problem. In such cases, the maximum cover committee can be computed exactly as we have done. There may be many elections where this is true and where the maximum cover committee can be computed without undue effort. For example, one could consider electing both the senators from a US state\footnote[2]{This would require amending the class system where each senator from a state is elected for different but overlapping terms.} from a set of aspirants through state-wide approval voting, and choosing the maximum cover committee.   

It is known that the greedy algorithm is often a good approximation to the solution of the maximum coverage problem.~\cite{coverage_proof}. Further, the greedy algorithm is computationally simpler than calculating the Hamming distance vector for each of the possible committees, and minimizing the $p$-norm. Therefore, the greedy algorithm applied to the problem of ensuring that the maximum number of voters get at least one of their approved candidates on the committee can be a computationally efficient, as well as reasonable, voting method, if the committee size $k$ is so large that the maxiumum cover problem cannot be solved exactly.
Applying the greedy algorithm to the example in Table~\ref{tab:main example}, the first candidate we choose is $ A_1, $ since we can cover 61\% of the voters with this. After choosing $ A_1, $ we can cover the most number of additional voters by choosing $ B_1, $ yielding the committee $ \{A_1, B_1\}$. This is the same result we get with the $p$-norm minimization method, with $ p = 2 $ and $  p = 3. $

\subsection*{The case where \boldmath$ p \to 0 $}

We now investigate the effect of letting $ p \to 0. $ In this case, it is the number of votes that the committee as a whole wins that matters, and not the number of votes of each of the constituent members. This can be shown using the same technique we used in the case of  $ p \to \infty. $  Using the same notation as before, we get 
$$
\lim_{p \to 0} \frac{||\mathcal{C}||_p}{||\mathcal{C^\prime}||_p} 
= \lim_{p \to 0} \frac{\left( \sum\limits_{i = 0}^{2k} \nu_i \cdot i^p  \right)^{1/p}}{\left( \sum\limits_{i = 0}^{2k} \nu^\prime_i \cdot i^p  \right)^{1/p} }
= \left(\frac{m - \nu_0}{m-\nu^\prime_0} \right)^{1/p}.
$$

So this method chooses the committee with the least value of $m- \nu_0$ or the committee with the highest value of $\nu_0$, namely approvals.
In the previous example from Table~\ref{tab:main example}, the result with $ p = 1 $ coincides with the one with $ p \to 0 $; we will therefore use a different example which is shown in Table~\ref{tab:p=0 example}. 

\begin{table}[ht]
	\begin{center}
		\begin{tabular}{|c|c|}
			\hline
			Candidate pair & Votes \\ \hline
			$ \{A_1, A_2\} $ &  300  \\ \hline
			$ \{A_1, B_1\} $ &  250  \\ \hline
			$ \{A_1, B_2\} $ &  150  \\ \hline
			$ \{A_2, B_1\} $ &  40   \\ \hline
			$ \{A_2, B_2\} $ &  60   \\ \hline
			$ \{B_1, B_2\} $ &  200  \\ \hline
		\end{tabular} 
		\caption{An example of a ballot to demonstrate the result of $ p \to 0 $.}
		\label{tab:p=0 example}
	\end{center}
\end{table}

We can see that the candidates with the most votes are $ A_1 $ and $  B_1, $ as shown in Table~\ref{tab:votes per candidate p=0}, and according to the minisum method, i.e., with $ p = 1,  $ the committee that will be elected is $ \{A_1, B_1\}$. However, reducing the value of $p$, we see, from Table~\ref{tab:p=0 distances} that the elected committee changes to $\{A_1, A_2\}$, the pair that obtained the most votes, even though candidate $ A_2 $ received the fewest number of votes in total. \footnotetext[3]{Strictly speaking, this should not be termed a norm for $p < 1$ since the triangle inequality is reversed.}

\begin{table}[ht]
	\begin{center}
		\begin{tabular}{|c|l|}
			\hline
			Candidate  & Number of approvals  \\ \hline
			$ A_1 $    & 300 + 250 + 150  = 700 \\ \hline
			$ A_2 $    & 300 + 40 + 60 = 400  \\ \hline
			$ B_1 $    & 250 + 40 + 200 = 490 \\ \hline
			$ B_2 $    & 150 + 60 + 200 = 410 \\ \hline
		\end{tabular} 
		\caption{The number of approvals per candidate, using the ballots from Table~\ref{tab:p=0 example}.}
		\label{tab:votes per candidate p=0}
	\end{center}
\end{table}

\begin{table}[ht]
	\begin{center}
		\begin{tabular}{|c|c|c|c|c|}
			\hline
			               &         \multicolumn{4}{c|}{Value of $ p $ in the $ p $-norm}         \\ \hline
			  Committee    &       1       &       0.5        &       0.1        &      0.001      \\ \hline
			$ \{A_1, A_2\} $ &     1800      &     1107.11      & \textbf{765.63 } & \textbf{700.62} \\ \hline
			$ \{A_1, B_1\} $ & \textbf{1620} & \textbf{1095.81} &      808.45      &     750.56      \\ \hline
			$ \{A_1, B_2\} $ &     1780      &     1225.51      &      914.08      &     850.62      \\ \hline
			$ \{A_2, B_1\} $ &     2220      &     1445.51      &     1040.44      &     960.77      \\ \hline
			$ \{A_2, B_2\} $ &     2380      &     1475.81      &     1026.70      &     940.83      \\ \hline
			$ \{B_1, B_2\} $ &     2200      &     1307.11      &      880.50      &     800.76      \\ \hline
		\end{tabular} 
		\caption{$ p $-norm\protect\footnotemark[3] of the $ m $-vector, raised to the power $p$, for all the possible committees, for different values of $ p, $ using the ballots from Table~\ref{tab:p=0 example}. We raise the $p$-norm to the power $p$ to keep the values bounded for small $p$. The value in bold is the minimizer for its column.}
		\label{tab:p=0 distances}
	\end{center}
\end{table}

\section{Ternary Voting}

It has not escaped our notice that this method of choosing the committee that minimizes the $ p $-norm of the vector of Hamming distances to the ballots, can be generalized to include ternary voting~\cite{ternary}. In this form of election, the voters can express their opinion about a fixed number of candidates, each of whom can either be approved (represented by a $1$) or rejected (represented by a $ -1 $). Consider the example from Table~\ref{tab:all preferences} and let the voters be allowed two opinions or non-zero entries. The voters who have preference vector $ (A, B, C) $ have three ways to express their preference using ternary voting, namely, $(1, 1, 0)$, $(1, 0, -1)$, or $(0, -1, -1)$.

In the case of only three representatives, the voters can actually indicate their full preference vector through ternary voting, but with more candidates, the full preference list is too complex to express. Moreover, it may be that the voter is actually neutral about most of the candidates, but rejects a few, and this form of ternary voting allows them to express this. 

Under ternary voting, we need to replace the Hamming distance between a candidate committee, $(c_1, c_2, \ldots, c_n)$ and a ballot, $(b_1, b_2, \ldots,b_n)$ by some suitable metric. Since the Hamming metric corresponds to the $ L_1 $-metric, namely, $\sum_{i=1}^n \lvert c_i - b_i \rvert$, it is natural to consider the same metric under ternary voting. 

However, we must revise our vector representation of the committees.  If we represent the hypothetical committee consisting of the first two out of four candidates as $(1,1,0,0),$ then the distance to the ballot $ (1,1,0,0) $ will be 0, whereas the distance to the ballot $(0,0,-1,-1)$ will be 4. This is a bad model since a voter disapproving of the last two candidates is implicitly approving the first two. Therefore, this candidate committee is better represented by $(1,1,-1,-1)$ in which case the distance to both the ballots will be 2, in accordance with our intuition.

Our generalization of approval voting from binary to ternary voting, to elect a committee of size $k$ from $n$ candidates, is the following. We represent each of the possible committees as a ternary vector $(c_1, c_2, \ldots, c_n)$ with $c_i = 1$ if candidate $i$ is in the committee, and $c_i = -1$, otherwise. Each of the $m$ approval ballots is a ternary vector $(b_1, b_2,\ldots, b_n)$ where $b_i = 1$ if the voter approves candidate $i$, $b_i = -1$ if the voter rejects candidate $i$, and $b_i = 0$ otherwise (the voter has no opinion). Each voter must express $k$ opinions and therefore make $k$ entries in his ballot as $1$ or $-1$; the rest of the entries must be zero. For the sake of mathematical simplicity (or elegance!), we have arbitrarily chosen to weigh approvals and rejections equally---a voter uses up one choice to approve a candidate, or to reject a candidate. 

It is interesting to ask what the properties of the $ p $-norm committees are, for various values of $ p $, under ternary voting.  Suppose candidate $i$ has $a_i$ approvals, $r_i$ rejections, and $ n_i = m - a_i - r_i $ neutral votes. Then, candidate $ i $ contributes \[ n_i + 2 r_i = m - a_i - r_i + 2 r_i = m - (a_i - r_i)\] to the  $ 1 $-norm if he is in the committee, and \[n_i + 2 a_i = m- a_i - r_i +2 a_i = m + (a_i - r_i)\] if he is not in the committee. Therefore, the minimum $ 1 $-norm committee consists of the $k$ candidates with the $ k $ least values of $ m - (a_i - r_i) $, or equivalently, the $ k $ highest values of $ a_i - r_i $. This is an intuitively pleasing extension of the minisum method to ternary voting: the $1$-norm committee minimizes the sum of the \emph{net} approvals of its members, when rejections (explicit disapprovals) are allowed. 

Similarly, we can extend our result of maximum coverage to ternary voting as well. Suppose we wish to elect a committee of $ k $ members from a set of $ n. $ Let each voter choose $ k $ candidates to either approve or reject. We see that the $ L_1 $-distance from the committees to the ballots ranges from $ n - k $ to $ n + k, $ and using the same method as before, we see that as $ p \to \infty, $ the committee that is selected is the one with the least number of ballots at distance $ n + k .$ It is clear that a ballot is only at this distance when none of the candidates that are approved are on the committee, \emph{and} all the candidates that are rejected \emph{are} on the committee. Therefore, in the context of the Maximum Coverage Problem, we consider a voter to be covered if any of the candidates he approves is on the committee, \emph{or} if any of the candidates he rejects is \emph{not} on the committee. Then, in this case also, choosing the committee that covers the maximum number of voters is better than using the minimax method because of the preponderance of ties that arise using the latter method.

We conclude this section with an example. Consider the example of Table~\ref{tab:main example} but with one change: the 500 voters who approved $\{A_1,A_2\}$ now choose to approve $A_1$ and reject $B_1$. Perhaps $B_1$ evokes strong emotions and the votaries of $A_1$ and $A_2$ would prefer to sacrifice approving $A_2$ and instead reject $B_1$.  All other votes remain the same. The results of this election are shown in Table~\ref{tab:ternary example} for various values of the $p$-norm.

\begin{table}[hb]
	\begin{center}
		\begin{tabular}{|*{7}{c|}}
			\hline
			&                      \multicolumn{6}{c|}{Value of $ p $ in the $ p $-norm}                       \\ \hline
			Committee    &       1       &       2        &       3        &       4        &      10       &      100      \\ \hline
			$ \{A_1, A_2\} $ &     3700 &     130.38      &     44.68      &     26.59      &     10.79      &     6.36      \\ \hline
			$ \{A_1, B_1\} $ &     3840      &     123.29 &     39.46 &     22.42      &     8.57      &     6.18      \\ \hline
			$ \{A_1, B_2\} $ & \textbf{3020} & \textbf{101.39} & \textbf{33.76} & \textbf{19.82} & \textbf{8.38} & \textbf{6.18}      \\ \hline
			$ \{A_2, B_1\} $ &     4980      &     161.12      &     51.97      &  29.73 & 11.21 & 6.39 \\ \hline
			$ \{A_2, B_2\} $ &     4160      &     133.27      &     42.74      &     24.41      &     9.65      &     6.28      \\ \hline
			$ \{B_1, B_2\} $ &     4300      &     147.65      &     49.38      &     28.84      &     11.18      &     6.38      \\ \hline
		\end{tabular} 
		\caption{$ p $-norm of the $ m $-vector for all the possible committees for different values of $p$, under ternary voting, using the ballots from Table~\ref{tab:main example} except that the 500 voters who approved $A_1$ and $A_2$ now choose to approve $A_1$ and reject $B_1$. The value in bold is the minimizer for its column.}
		\label{tab:ternary example}
	\end{center}
\end{table}

For every value of $p$, the committee elected is $\{A_1,B_2\}$. Compare this result with Table~\ref{tab:p norm distances} where this particular committee is not elected for any value of $p$. Moreover, $B_1$ was elected in the binary case for all $p\ge 2$ but is never elected in this ternary case. The option to reject a candidate, as opposed to approve another candidate, can dramatically change the result of an election!

\section{Summary}

We see that all three known approval voting methods---the committee with the most approvals, the minisum committee and the minimax committee---are all special cases of our $p$-norm method for $ p \to 0,$ $p = 1 $ and $ p \to \infty $ respectively. However, using an intermediate value of $p$ such as 2 or 3 is probably a better choice than any of these methods as a compromise between approvals and disapprovals. 

The minimax method is particularly unsuitable for this type of election given the preponderance of ties, especially when the number of voters is large, and should be logically replaced by the Maximum Cover method, which too is a special case of our $p$-norm method for large, but finite, $p$, and intuitively corresponds to minimizing disapprovals.

We also showed that our $p$-norm method can be generalized to the problem of ternary voting, and the interpretations of the results when $ p = 1 $ and $ p \to \infty $ are natural analogs of the binary case. 

\section*{Acknowledgement}

I would like to thank Professor Y. Narahari for giving me the opportunity to work on this summer project, and Palash Dey for his guidance. I would also like to acknowledge the KVPY program at IISc for enabling me to pursue a research-oriented undergraduate program. I would like to thank the anonymous referee for his or her insightful review, and both the referee and the editor for their thoughtful suggestions that have improved this paper. 

%\nocite{*}
\bibliographystyle{siam}  \bibliography{./biblio}

\end{document}